# Turbo Spin Echo Imaging at 7T with Bilateral Orthogonality Generative Acquisitions Method for Homogeneous $T_1$, $T_2$ and Proton Density Contrasts

Çelik Boğa and Anke Henning

**Abstract:**

**Purpose:** Bilateral Orthogonality Generative Acquisitions (BOGA) method, which was initially implemented for $T_2^*$ contrast via gradient echo acquisitions, is adapted for TSE imaging at 7T using parallel transmission (pTx) system for obtaining homogeneous $T_1$, $T_2$ and proton density weighted images.

**Theory and Methods:** Multiple TSE images with complimentary RF modes and scan parameters are acquired as input images for the BOGA method where RF modes have complimentary transmit and receive field inhomogeneity patterns and scan parameters have varying echo and repetition times. With the application of the BOGA method using different subsets of the data acquisitions for each contrast, homogeneous $T_1$, $T_2$ and proton density contrast in the final images obtained. Furthermore, to demonstrate the effect of the TSE factor, two TSE factors are used individually. Normalized intensity profiles and signal to noise ratio maps are utilized for the comparison of the CP mode images and the TSE factors respectively.

**Results:** Homogeneous $T_1$, $T_2$ and proton density weighted images are obtained with the TSE implementation of the BOGA method without the transmit and receive field inhomogeneity effects. Furthermore, mixed contrast effects of the TSE acquisition are simultaneously resolved independently of the TSE factor.

**Conclusion:** TSE application of BOGA method results in homogeneous $T_1$, $T_2$ and proton density contrasts at 7T, as the inhomogeneity effects are removed from the final contrast without any prior data acquisitions.

## Introduction:

Rapid Acquisition with Relaxation Enhancement (RARE) has been introduced for accelerating spin echo acquisitions by utilizing multiple refocusing pulses to read multiple k-space lines with a single excitation[1]. Commercial implementations of this technique are named Turbo Spin Echo or Fast Spin Echo depending on the vendor[2]. In this work, utilization of the RARE technique will be referred as Turbo Spin Echo (TSE) imaging.

Turbo Spin Echo imaging has been widely utilized in clinical applications[3-8] due to its versatility and robustness against geometric distortions caused by the main field ($B_0$) inhomogeneity (Δ)[1] in 1.5T and 3T MRI systems. Many applications such as brain stem[3,4], spinal cord[5,6] and musculoskeletal[7,8] imaging, are heavily relying on the utilization of TSE) sequences. While TSE imaging is commonly utilized in 1.5T and 3T MRI systems, few applications have been demonstrated at Ultra High Field MRI systems[9-11] because of the increased specific absorption rate (SAR) and sensitivity to transmit field ($B_1^+$) inhomogeneity of the refocusing pulses[12].

Due to the higher spatial resolution and signal to noise ratio, interest in 7T human magnetic resonance imaging (MRI) systems has increased, especially in neuroimaging research as well as diagnostic imaging and patient management in brain and musculoskeletal disorders[13-15]. However, transmit field ($B_1^+$) and static magnetic field ($B_0$) inhomogeneity (Δ) at 7T results in loss of signal and unexpected changes in the contrast, which can result in inaccuracies of the diagnosis[12,16] and difficulties in interpretation of research results. It has been previously shown that these issues can limit the utilization of 7T human

MRI systems in general for clinical diagnostics and especially of TSE imaging calling for additional technical considerations[17,18]. In literature, methods for mitigating the $B_1^+$ inhomogeneity using parallel transmission (pTx) systems have been introduced[19-34].

One of the previously proposed methods is the Time Interleaved Acquisitions of Modes (TIAMO) method, where two different radiofrequency (RF) modes are acquired in an interleaved fashion and combined as sum of squares to mitigate the signal dropouts due to $B_1^+$ inhomogeneity[19]. In the initial implementation of TIAMO, two RF modes are utilized: (i) Circularly polarized (CP) mode where subsequent channels have $45^o$ phase difference and (ii) "Gradient mode" where subsequent channels have $90^o$ phase difference in an 8 channel pTx system[19]. The TIAMO method was then further developed to utilize designed RF modes using a priori $B_1^+$ shimming to obtain better homogeneity at the target region in TSE imaging[20]. While the TIAMO method largely reduces signal dropouts due to the $B_1^+$ inhomogeneity, resulting images still have remaining spatially varying signal intensities and image contrast, especially towards the peripheral regions due to remaining transmit inhomogeneity and inhomogeneous receive sensitivity[19-22].

Another method for addressing the $B_1^+$ inhomogeneity at 7T is the parallel transmission (pTx) pulse design, where a radiofrequency (RF) pulse is designed for each channel of the pTx system using the prior measurements of $B_1^+$ and $\Delta B_0$[23-28]. However, these measurements of $B_1^+$ and $\Delta B_0$ can prolong the scan time and require additional time for the design of the pulses[23-28]. Single channel $B_1^+$ measurement protocols as needed as input for pTx RF pulse design can also lead to measurement errors due to the large range of flip angles[29]. In order to circumvent the requirement for the data acquisition before each scan, universal pTx pulses have been introduced[30-33] where pTx pulses are designed over a database of $B_1^+$ and $\Delta B_0$ measurements, thus removing the requirement for additional scan can computation time during the scan session. However, one drawback of this method is that a library of pTx pulses must be designed individually for individual RF coil designs and target body part depending on the application. Moreover, in the previously demonstrated applications with TSE, resulting images are still susceptible to the receive channel inhomogeneity[32,33]. Recent fast on-scanner customized pTx pulse design methods increase the total scan time by approximately 67 seconds[34], to improve the performance of the universal pulses. However, for reducing the required scan time for the performance improvement, $B_1^+$ and $\Delta B_0$ measurements are acquired with lower resolution than the target resolution[34].

In literature, it has been shown that an increased length of the echo train in TSE imaging, the so called TSE factor, results in a mixed image contrast in the final images as $T_2$ decay becomes prominent through the later acquisitions in the echo train[35]. This imposes a limit on the realizable acceleration using high TSE factors without compromising a predefined image contrast, as faster scans might not result in useful and consistent contrast. As $T_2$ decreases with the increase of the main field strength[36,37], TSE applications at 7T is critically affected from this issue, particularly $T_1$ and proton density (PD) contrasts as $T_2$ weighting through the end of echo train can be more prominent than the $T_1$ and PD contrasts.

In this work, the previously introduced Bilateral Orthogonality Generative Acquisitions (BOGA) method[38-40] which was implemented for homogeneous whole brain $T_2^*$ contrast at 7T has been extended to multi-contrast TSE imaging for obtaining homogeneous images with clearly defined $T_1$, $T_2$ and PD contrasts that are free of $B_1$ (transmit ($B_1^+$) or receive ($B_1^-$)) inhomogeneities using dual channel 7T MRI system. For the application of the multi-contrast TSE BOGA method, 6 data acquisitions are utilized with 3 scan parameters (SPs) and 2 RF modes. SPs have varying echo times (TE) or repetition times (TR) to mimic the conventional TSE parameters for $T_1$, $T_2$ and PD contrasts. For each contrast, the BOGA method is implemented individually. On the other hand, RF mode refers to a set of $B_1$ inhomogeneity pattern in the acquired images which are chosen to have complimentary high and low intensity regions. For the implementation of the BOGA method for multi-contrast TSE imaging, the TSE factor must be the same for all six data acquisitions to keep the $B_1^+$ inhomogeneity effects consistent for all acquisitions. In contrast to the CP mode images the aforementioned non-consistent image contrast of conventional TSE imaging is inherently addressed with the BOGA method. Since the TSE factor is the same for all input images the common additional $T_2$ decay due to the TSE acquisitions will be mitigated by the BOGA method similar to the $B_1^+$ inhomogeneity effects. Moreover, the comparison between the final images with $T_1$, $T_2$ and PD contrasts with two TSE factors are also presented. This paper presents the framework for the extension of the BOGA method to other contrasts than $T_2^*$ contrast.

## Theory:
### Bilateral Orthogonality Generative Acquisitions (BOGA) Method

The TSE implementation of the BOGA method in a dual channel pTX system[38-40], requires 2 complimentary RF modes resulting in non-overlapping low intensity regions and 2 distinct sets of scan parameters (SP) to obtain images using (i) the first RF mode and the first SP, (ii) the second RF mode and the first SP, (iii) the first RF mode with the second SP and (iv) the second RF mode with the second SP. As the image contrast of the final image depends on the difference between TE and TR of the SPs, the BOGA method must be implemented individually for each contrast. For three image contrasts - $T_1$, $T_2$ and PD weighted imaging - a total of 6 different acquisitions are needed. Subset of these images will be combined per desired image contrast as detailed below.

The first and second input images of the BOGA method can be modeled as $S_1 = A_1(g_1 + g_2)$ and $S_2 = A_1(-g_1 + g_2)$, whereas the third and fourth input images are calculated as $S_3 = 0.5(-S_3^{pre} + S_4^{pre})$ and $S_4 = 0.5(S_3^{pre} + S_4^{pre})$. $S_3^{pre}$ and $S_4^{pre}$ are the images described in (iii) and (iv) and modeled as $S_3^{pre} = A_2(g_1 + g_2)$ and $S_4^{pre} = A_2(-g_1 + g_2)$), respectively. Here, $A_i$ is the weighting of the image due to the TE and TR of the SP set $i$, and $g_{1,2}$ are the observed transmit channels that result in the different RF modes used in data acquisition.

$B_1$ inhomogeneity pattern of the observed single transmit channels, $S_3$ and $S_4$, do not necessarily coincide with the actual single transmit channels' $B_1$ inhomogeneity pattern, as $S_3$ is calculated using the summation of utilized RF modes and $S_4$ is calculated using the difference of the RF modes. Whereas the $B_1$ inhomogeneity pattern of the actual transmit channel are independent of the chosen RF modes. In TSE applications, inhomogeneity pattern of the observed channels are also dependent on the TSE factor along with the initial selection of RF modes, as a cascade of inhomogeneous RF pulses, in combination with the $B_1^-$ inhomogeneities can result in different $B_1$ inhomogeneity patterns than the initial RF mode. Overall $B_1$ inhomogeneity effects are included in the calculated observed transmit channels, $S_3$ and $S_4$, and are mitigated by the application of the BOGA method.

Then, four input images ($S_1$, $S_2$, $S_3$ and $S_4$) are combined into a first set of two intermediate images $C_1 = S_3^* S_1 + S_4 S_2^*$ and $C_2 = S_4^* S_1 - S_3 S_2^*$ as described in more detail previously[38,40]. A complimentary set of two intermediate images are calculated as $D_1 = S_3^* S_1 - S_4 S_2^*$ and $D_2 = S_4^* S_1 + S_3 S_2^*$. Then, the final images are obtained as $I = 0.25\,(C_1 + D_1 + C_2 + D_2 + (C_1 - D_1 + C_2 - D_2)^*)/(|S_3|^2 + |S_4|^2)$[39,40]. The result is a homogeneous image where magnitude and phase of the image is defined by the contrast and phase difference arising from the difference of the SPs. As the phase difference due to the SPs are not of interest herein, the focus of this work will be on magnitude images.

Since the final image contrast is determined by the difference between TE and TR of SPs, three SPs are utilized as SP$_1$ (short TE and short TR), SP$_2$ (short TE and long TR) and SP$_3$ (long TE and long TR). The $T_1$ contrast is achieved by selecting SP$_2$ as the first SP for $S_1$ and $S_2$, and SP$_1$ as the second SP for $S_3$ and $S_4$, as the final contrast is due to the difference between TRs of SP$_1$ and SP$_2$. Whereas the $T_2$ contrast is achieved by selecting SPs with different TEs, namely SP$_3$ as the first SP for $S_1$ and $S_2$, and SP$_2$ as the second SP for $S_3$ and $S_4$. However, for the realization of the PD contrast, the average of SP$_1$ and SP$_3$ is used as the second SP for $S_3$ and $S_4$, and SP$_2$ is selected as the first SP for $S_1$ and $S_2$. Averaging of SP$_1$ and SP$_3$ results in an image with a contrast corresponding to longer TE and shorter TR than SP$_3$, which creates the PD contrast in the final image obtained via BOGA method.

# Methods:

## Scan Parameters

In this work, two healthy volunteers (1 male (27) and 1 female (55)) were recruited. Signed consent for the scans was acquired for the study, in compliance with the institutional review boards' conditions for in vivo studies. Data was acquired with a 7T Philips Healthcare DSync human MRI system using a two-channel transmit head coil (Nova Medical) along with a 32 channel receive coil insert (Nova Medical). For each healthy volunteer, two sets of 3D TSE images with TSE factors 50 and 100 are acquired. Each set of TSE images includes a total of 9 input images with each RF mode (CP mode, RF mode 1 and 2) and SP. Only TSE images obtained via RF mode 1 and RF mode 2 are utilized for the BOGA method and compared against the CP mode images. $T_1$, $T_2$ and PD weighted images for each RF mode acquired with both TSE factors are utilized for demonstrating the robustness of the BOGA method to the effects of an increasing TSE factor. The performance of the BOGA method for the mitigation of the $B_1^+$ inhomogeneity is demonstrated within each individual data set with identical TSE factor.

For each 3D TSE acquisition, with $90^0$ flip angle and $60^o$ refocusing angle with refocusing control, an acquisition voxel size of 1x1x1mm, field of view of 180x220x170, 1 average and Compressed SENSE factor of 8. TEs, and TRs for SPs for each TSE factor are selected with respect to the TSE factor (Table 1). However, the difference between TEs and TRs are kept the same for each TSE factor to generate identical image contrasts for both TSE factors utilized herein. As the difference between TEs and TRs are constant between each TSE factor, final images obtained via the BOGA method are expected to have similar contrasts. TE, effective TE and TR values for each SP for both sets are presented in Table 1. Total scan time for all 3 TSE acquisitions with all SPs using a single RF mode for low TSE factor is 18 minutes 48 secs and for high TSE factor 13 minutes and 55 seconds. Total scan time for the BOGA method is 37 minutes 36 seconds for low TSE factor and 27 minutes 52 seconds for high TSE factor since both RF mode 1 and RF mode 2 is utilized.

|         | TSE factor 50 | | | TSE factor 100 | | |
|---------|-----|------|------|------|------|------|
|         | $SP_1$ | $SP_2$ | $SP_3$ | $SP_1$ | $SP_2$ | $SP_3$ |
| TE (ms) | 8 | 8 | 28 | 8 | 8 | 28 |
| TR (ms) | 1500 | 2750 | 2750 | 2600 | 3850 | 3850 |

**Table 1:** TE and TR values for each SP with respect to the used TSE factor. TE difference between SPs are chosen to be 20 ms and TR difference is chosen to be 1250 ms for both TSE factors

RF mode 1 and 2 are chosen as the individual channels of the dual channel pTx system as they have complimentary $B_1$ inhomogeneity patterns by design to obtain the CP mode (high intensity at the center and low intensity at the periphery) when used simultaneously. Another reason for the selection of individual transmit channels is to adhere to the specific absorption rate (SAR) limits as a train of RF pulses in a TSE sequence can result in high SAR values for an arbitrary RF shim. This also enables the implementation of the BOGA method within vendor specified SAR limits as transmit channel cycling is an option provided by the vendor.

Display masks are calculated for each volunteer based on the magnitude images of the CP mode high TSE factor data acquisition with $SP_2$. Then for each transversal slice, a mask is obtained by thresholding to remove pixels with an intensity below the 0.02% of the maximum intensity of that slice. Display masks are only applied for the visualization of the images obtained with BOGA method, they are not used for SNR calculations.

**Normalized Intensity Profiles**

For further visualization of the inhomogeneity effects in the images', normalized profiles are calculated by dividing each pixel with the average value of the pixel intensities on the profile. Rationale for the normalization is to remove pixel intensity scaling effects from inhomogeneity effects. Resulting profiles indicate the relative change of the values along the profile with respect to the average value of the pixels in the profile, while preserving the relative intensity changes between different tissues.

**Signal to Noise Ratio (SNR) Mapping**

A SNR map for each image is calculated using the SNR calculation method described in[41]. This method applies a 5x5x5 voxel spatial smoothening box filter to the image itself, and the resulting smoothened image is then subtracted from the initial image to obtain a noise image. Standard deviation for each pixel of the noise image is then calculated using a 5x5x5 moving kernel[42]. Then the actual SNR for each pixel is calculated by dividing the original signal intensity with the calculated standard deviation per pixel. SNR maps are used for comparison of the impact of the TSE factor on the SNR of the final images as well as the SNR comparison between CP mode images with the images obtained with the BOGA method. For eliminating the effect of the total scan time, CP mode SNRs were multiplied by $\sqrt{2}$ to achieve the theoretical SNR increase by 2 averages matching the scan time for BOGA method.

# Results:

In Figure 1, TSE images obtained using CP mode with all three SPs for each TSE factor are illustrated in transversal, coronal and sagittal orientations. Effects of $B_1$ inhomogeneity is prominent in every image as the center is significantly brighter than the periphery of the brain. It can also be observed that high TSE factor images contain prominent $T_2$ weighting independently of the SP, whereas images acquired with SP$_1$, SP$_2$ and SP$_3$ are expected to show $T_1$, PD and $T_2$ contrast respectively. This demonstrates the limitation of the realizable TSE factors for obtaining $T_1$ and PD contrasts with conventional TSE at 7T. Even with the low TSE factor (50), contrasts obtained via SP$_1$ and SP$_2$ do not represent clear $T_1$ and PD contrasts in the TSE images. Whereas TSE images obtained for the high TSE factor (100), shows dominant $T_2$ contrast for all three SPs. Low intensity at the left cerebellar lobe can be observed for all images as the coverage of the dual transmit channel coil is limited at the cerebellum region.

Transversal, coronal and sagittal slices of the 3D TSE images for both volunteers obtained with RF mode 1 and RF mode 2 with all SPs are demonstrated in Figure 2 and Figure 3 respectively, for high and low TSE factors. Complimentary nature of the RF mode is clearly observable for each volunteer as RF mode 1 results in images with lower intensity towards the frontal and posterior regions of the brain and higher intensity towards the left and right regions of the brain. Whereas RF mode 2 results in images with higher intensity towards the frontal and posterior regions of the brain and lower intensity towards the left and right regions of the brain. Both RF modes result in sufficient signal intensity at the center of the brain. Similarly to the regular operating mode of the 7T scanner, CP mode, both RF modes have insufficient coverage towards the left cerebellum region, where images obtained with both RF modes show significant loss of signal intensity.

Magnitudes of the input images ($S_1$, $S_2$, $S_3$ and $S_4$) of BOGA method that are calculated for $T_1$, $T_2$ and PD contrasts for each TSE factor are presented in Figure 4 in transversal orientation. $S_3$ images calculated from the difference of the images acquired with RF mode 1 and 2 have higher intensity at the peripheral regions, where one of the RF modes has low signal intensity and the other has high intensity. Whereas $S_4$ resembles the CP mode images illustrated in Figure 1 as it is calculated as the means of the images acquired with RF modes 1 and 2, which as expected since CP mode is obtained by the simultaneous use of both RF modes.

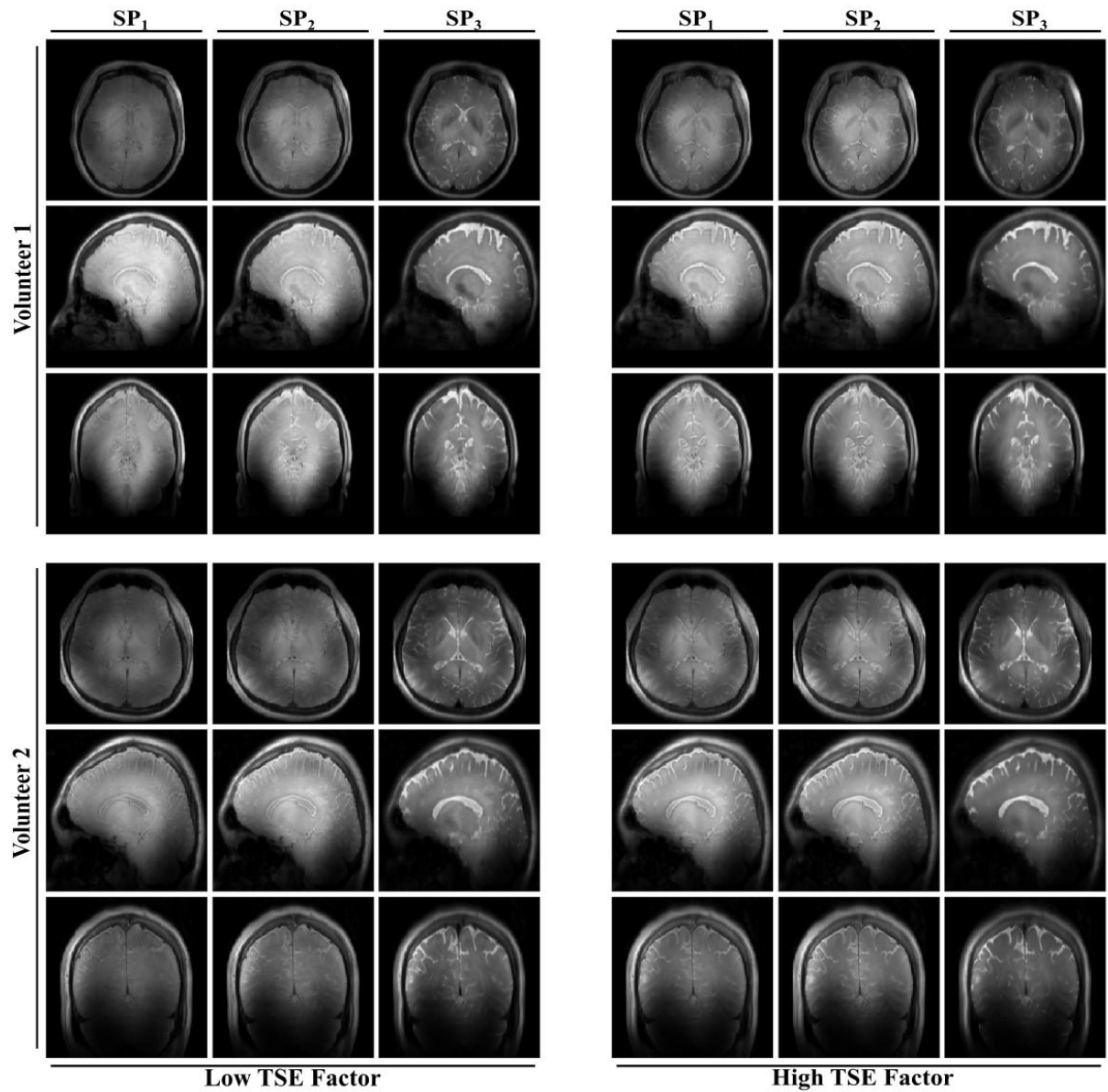

**Figure 1:** TSE images acquired in CP mode using $SP_1$, $SP_2$ and $SP_3$ and low TSE factor of 50 and high TSE factor of 100 for two volunteers. Images acquired with low TSE factor show mixed contrast whereas images obtained with high TSE factor show $T_2$ contrast. For both volunteers, effects of $B_1$ inhomogeneity can be observed as high intensity at the center and low intensity towards the periphery of the brain. Significant decrease in the signal intensity can be observed for the cerebellum region.

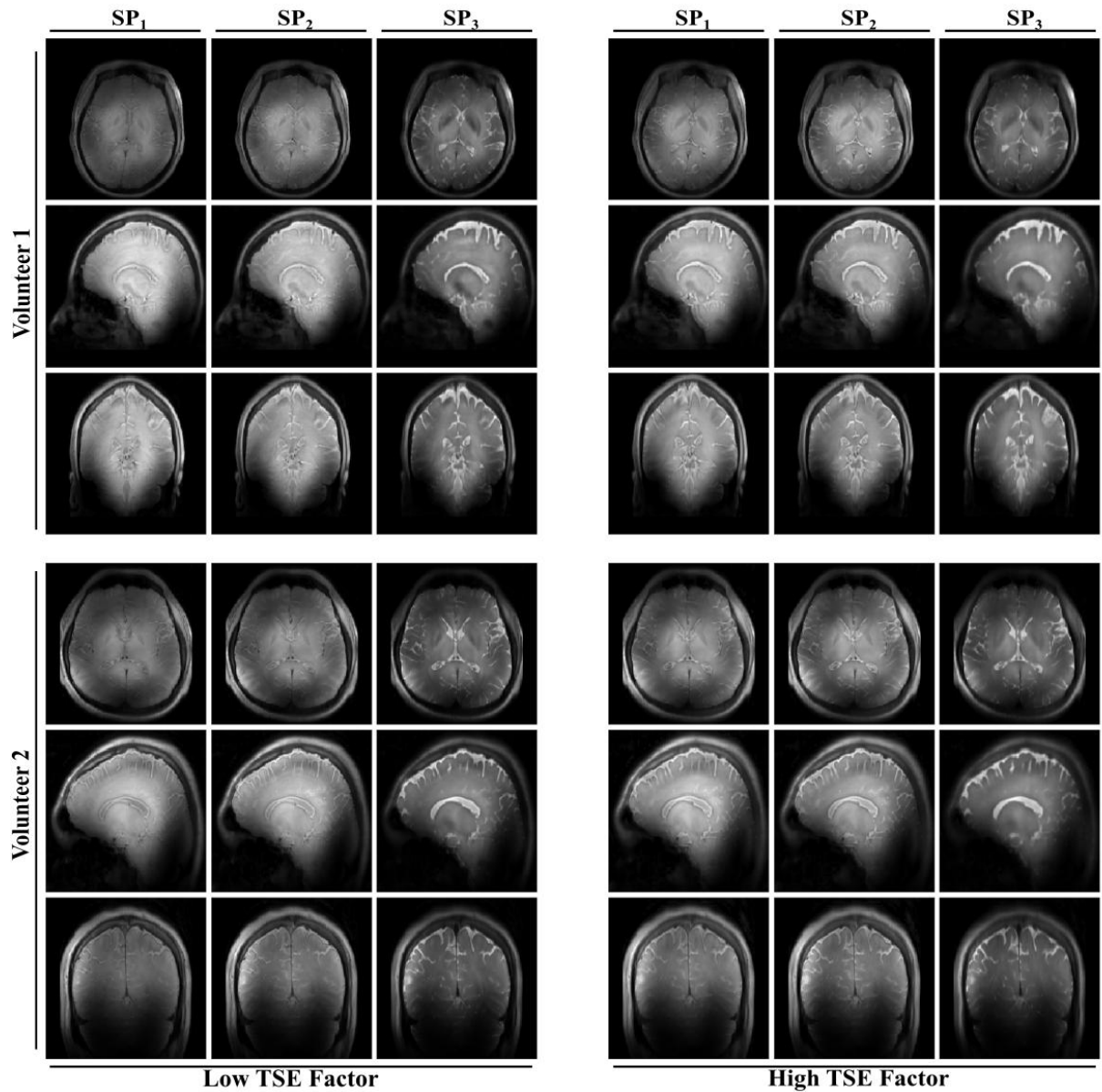

**Figure 2:** TSE images acquired in RF mode 1 using $SP_1$, $SP_2$ and $SP_3$ and low TSE factor of 50 and high TSE factor of 100 for two volunteers. Similarly to CP mode TSE images, low TSE factor results in mixed contrast and high TSE factor results in dominant $T_2$ contrast. Low intensity regions at the front can back of the brain are observed as the $B_1$ inhomogeneity pattern of the RF mode 1.

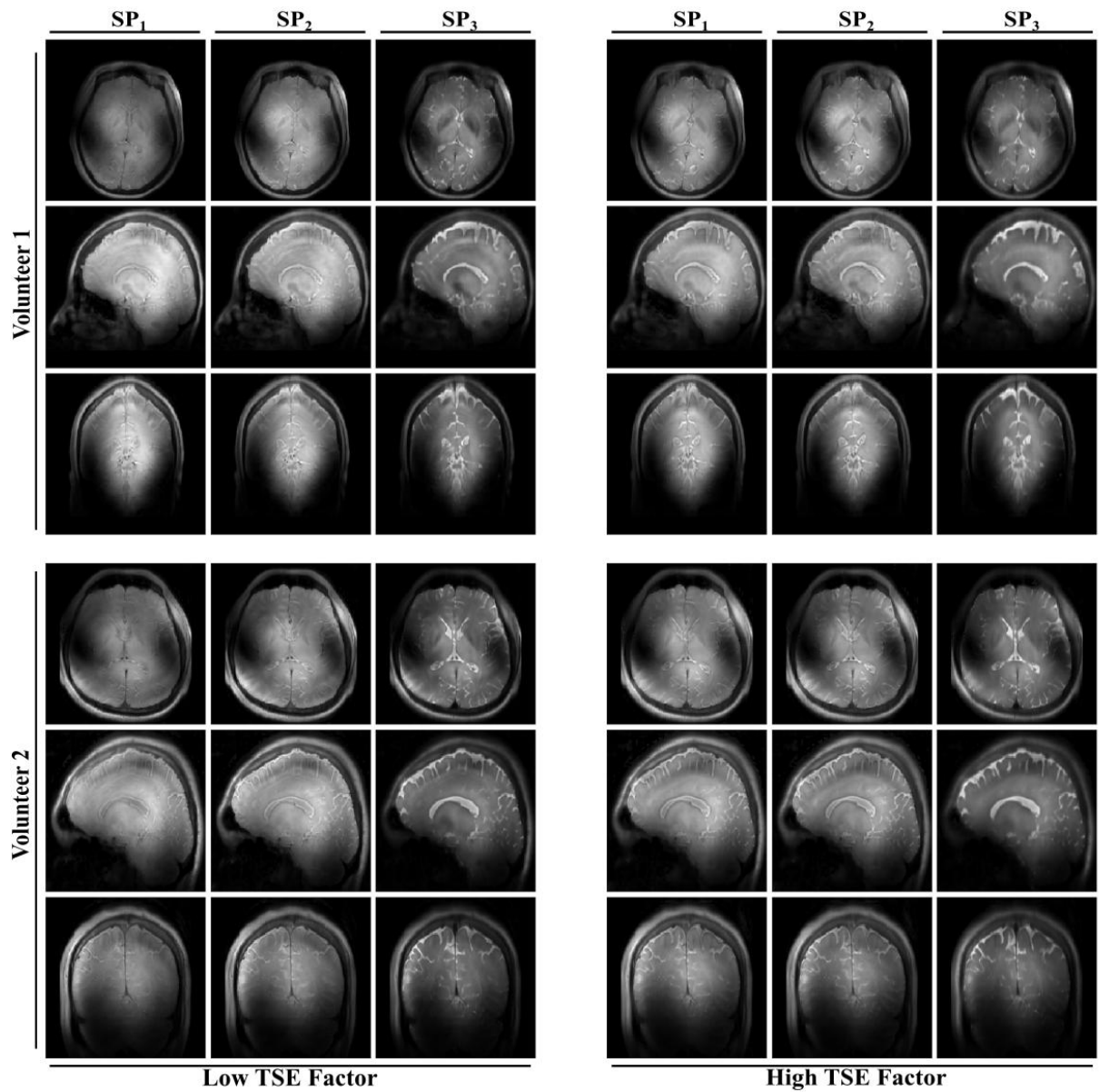

**Figure 3:** TSE images acquired in RF mode 2 using $SP_1$, $SP_2$ and $SP_3$ and low TSE factor of 50 and high TSE factor of 100 for two volunteers. In comparison to the RF mode 1, low intensity regions are shifted to the sides of the brain for RF mode 2, demonstrating the complimentary $B_1$ inhomogeneity patterns of the RF modes. However, for both, coil coverage in cerebellum is not sufficient.

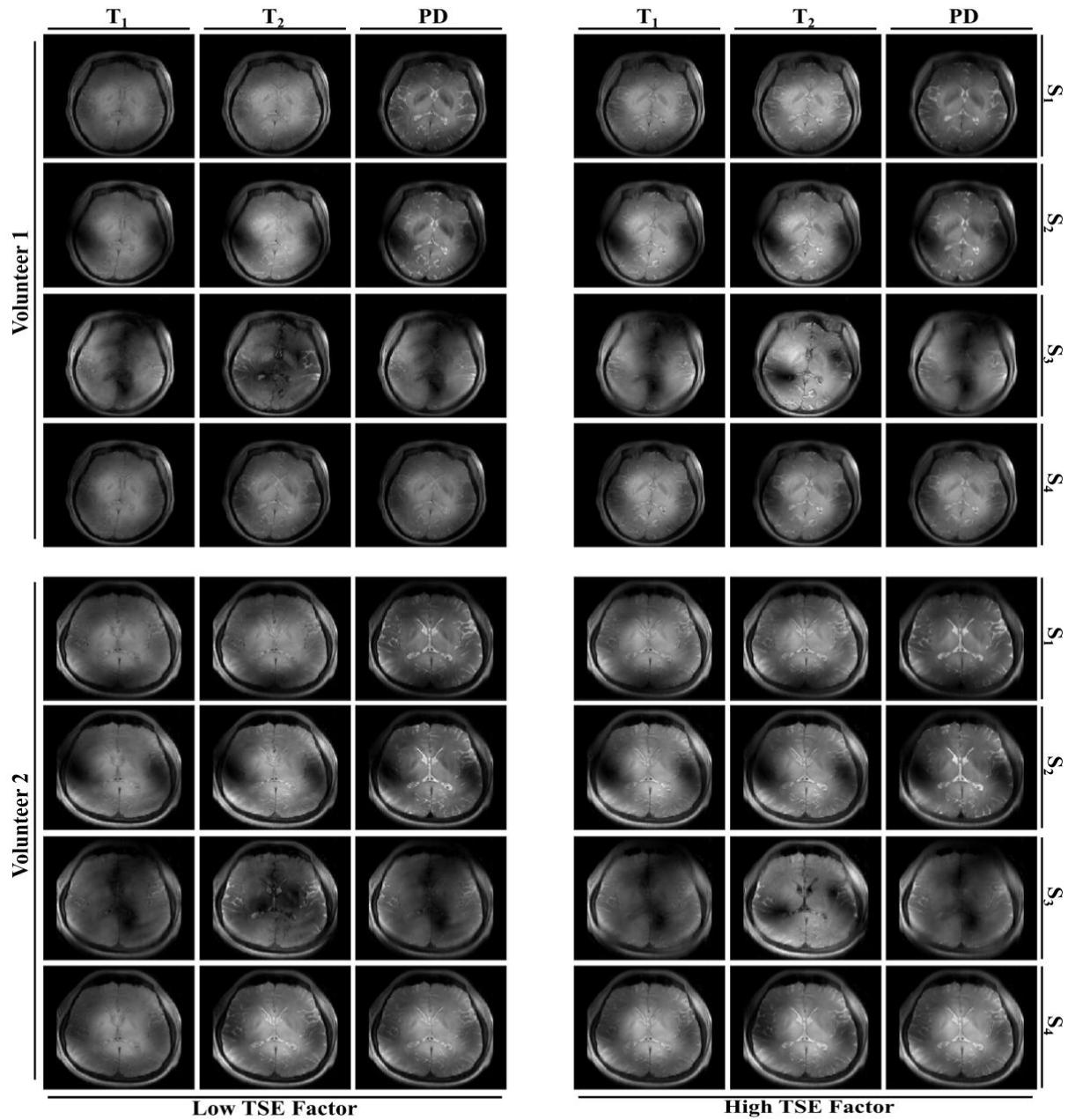

**Figure 4:** Magnitudes of the input images ($S_1$, $S_2$, $S_3$ and $S_4$) of BOGA method calculated for $T_1$, $T_2$ and PD contrasts for both TSE factors and volunteers. $S_3$ shows higher intensity at the peripheral regions where the $B_1$ inhomogeneity patterns of RF mode 1 and 2 differ considerably. On the contrary, $S_4$ shows higher intensity at the center where both RF modes have high intensity, similarly to CP mode $B_1$ inhomogeneity pattern.

Figure 5 illustrates the $T_1$ weighted images obtained via the BOGA method using the respective input images shown in Figure 4, in transversal, sagittal and coronal orientations for both TSE factors and volunteers. These $T_1$ weighted images are free of $B_1$ inhomogeneity effects and present a clear and homogenous image contrast unlike the TSE images shown in Figure 1. Compared to the CP mode images, $T_1$ weighted images obtained via the BOGA method do not suffer from variations in signal intensity and mixed contrast, indicating that the BOGA method addresses both issues – the $B_1^+$ inhomogeneity and the increasing $T_2$ weighting at long TSE factors, simultaneously. However, because of a lack of coil coverage, hyperintense regions in the left cerebellar lobe can be observed for the $T_1$ weighted images (indicated by red arrows), which cannot be addressed by the BOGA method as coil coverage limitations are common in every RF mode.

In Figure 6, $T_2$ weighted images obtained via the BOGA method are shown for both volunteers in transversal, sagittal and coronal orientations for each TSE factor. $T_2$ contrast in the images obtained via the BOGA method are in line with the $T_2$ contrast in the TSE images in Figures 1, 2 and 3 as expected. However, compared to the CP mode images in Figure 1, the effect of the $B_1$ inhomogeneity of the CP mode is mitigated in the $T_2$ weighted images obtained via BOGA method. These images also present a spatially more uniform $T_2$ contrast as the $T_2$ contrast for BOGA derived images depends on the difference of the TEs between the input images. In contrast, for conventional CP model TSE images the $T_2$ contrast depends on the weighted average of the $T_2$ decays in TSE train. In the left side of the cerebellum, an increase in the image intensity can be observed in the BOGA derived $T_2$ weighted images, in contrast to the $T_1$ weighted images, they are not as severely affected by the coil coverage issue. However, in combination with the increased $T_2$ decay in case of the high TSE factor, hyperintense regions in cerebellum (indicated by red arrows) in the $T_2$ images obtained can be observed for volunteer 2. These hyperintense regions are not present in the low TSE factor images.

PD weighted images for both volunteers obtained via the BOGA method using the respective input images shown in transversal, sagittal and coronal orientations for each TSE factor are demonstrated in Figure 7. Similarly to $T_1$ weighted images in Figure 5, PD weighted images are also free of $B_1$ inhomogeneity effects and present a clear and homogeneous PD contrast compared to the TSE images shown in Figure 1, 2 and 3. However, unlike the $T_1$ weighted images, coil coverage limitations present itself as hypointense regions in the cerebellum (indicated by red arrows), that exists in all three images as low intensity in this region occurs for both RF modes.

For both volunteers, homogeneous images with similar $T_1$, $T_2$ and PD contrasts can be obtained with both low and high TSE factors, as the contrast of the final images obtained by the BOGA method depends on the difference between SPs. This illustrates that the same image contrast can be achieved independently of the TSE factor with the BOGA method. This is a major improvement of BOGA TSE over conventional TSE imaging at 7T, which can result in mixed contrast depending on the TSE factor. However, the resulting range of $T_1$ and $T_2$ weighting is limited for the high TSE factor acquisitions, due to the increased TR and TSE factor.

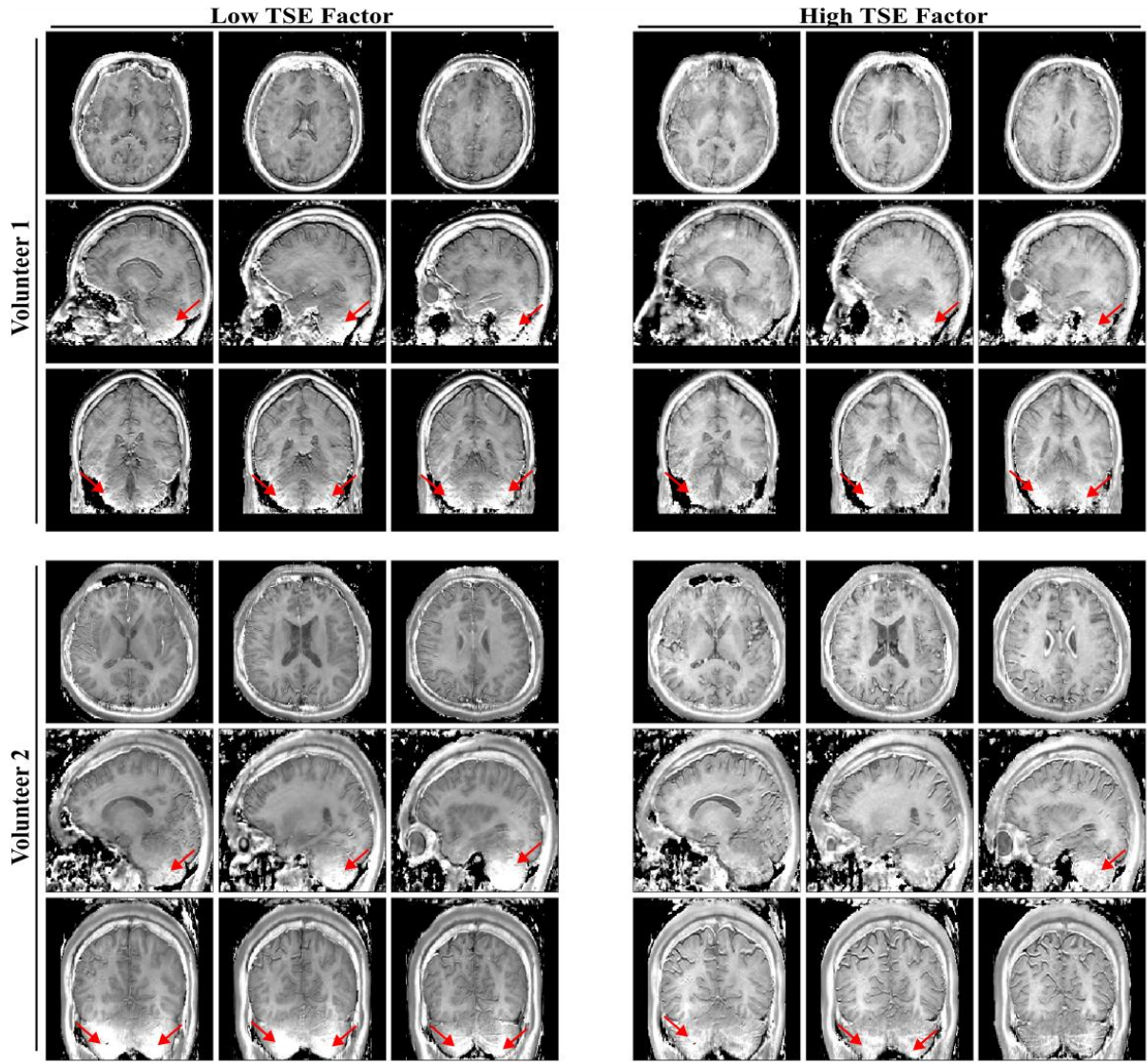

**Figure 5:** Homogeneous images with $T_1$ contrast obtained via BOGA method using TSE acquisitions with SP$_1$ and SP$_2$, in transversal, sagittal and coronal orientations. Bright regions can be observed at the cerebellum and brain stem regions where the RF coil coverage is not sufficient for either RF mode. Hyperintense bright regions are indicated with red arrows in sagittal and coronal slices.

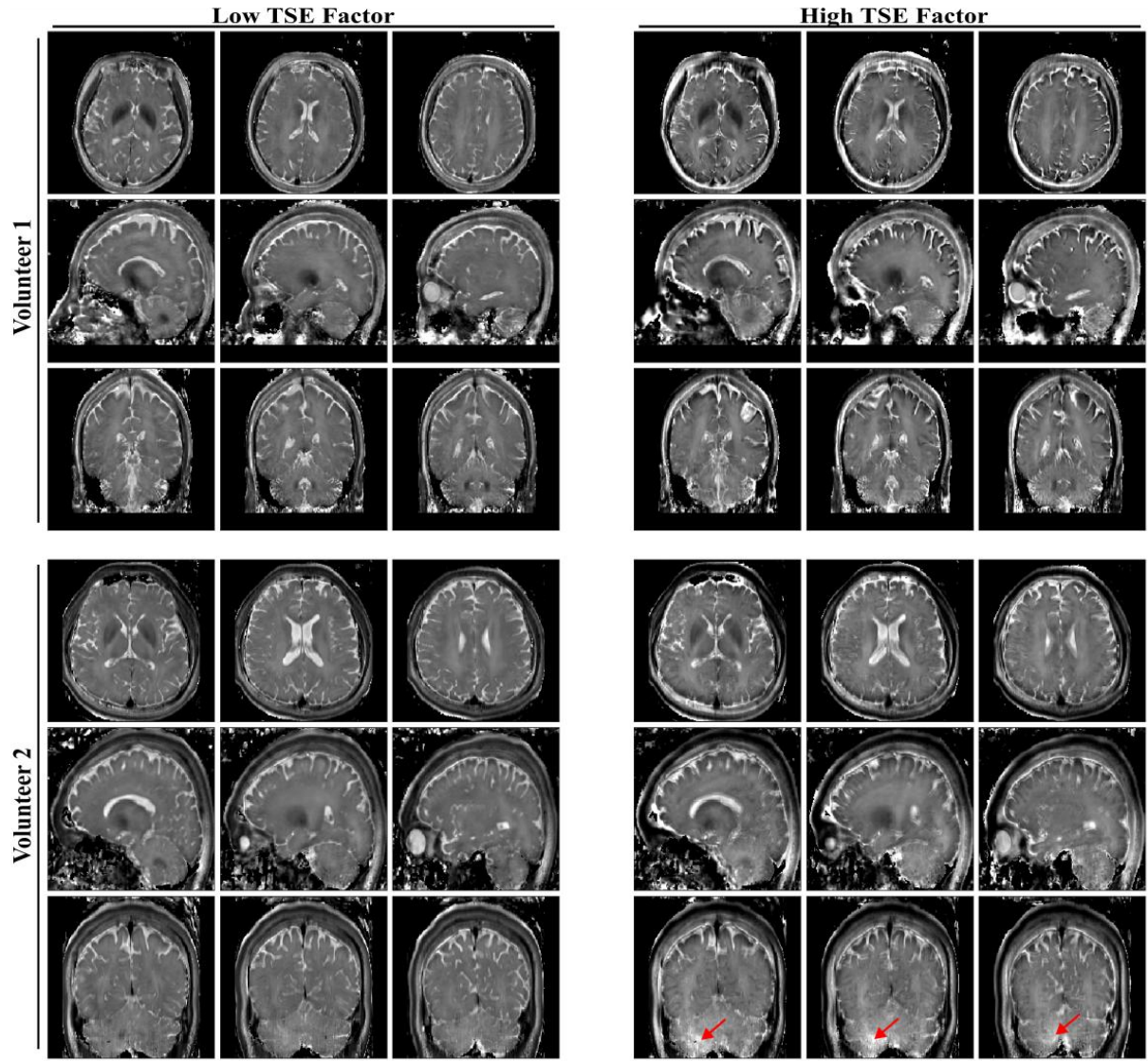

**Figure 6:** Homogeneous images with $T_2$ contrast obtained via BOGA method using TSE acquisitions with $SP_3$ and $SP_2$, in transversal, sagittal and coronal orientations. Stratum and cerebellum regions can be observed without the effect of $B_1$ inhomogeneity. Hyperintense regions in cerebellum region of volunteer 2, resulting from the combined effects of increased $T_2$ decay of the high TSE data acquisition and the coil coverage, are indicated with red arrows.

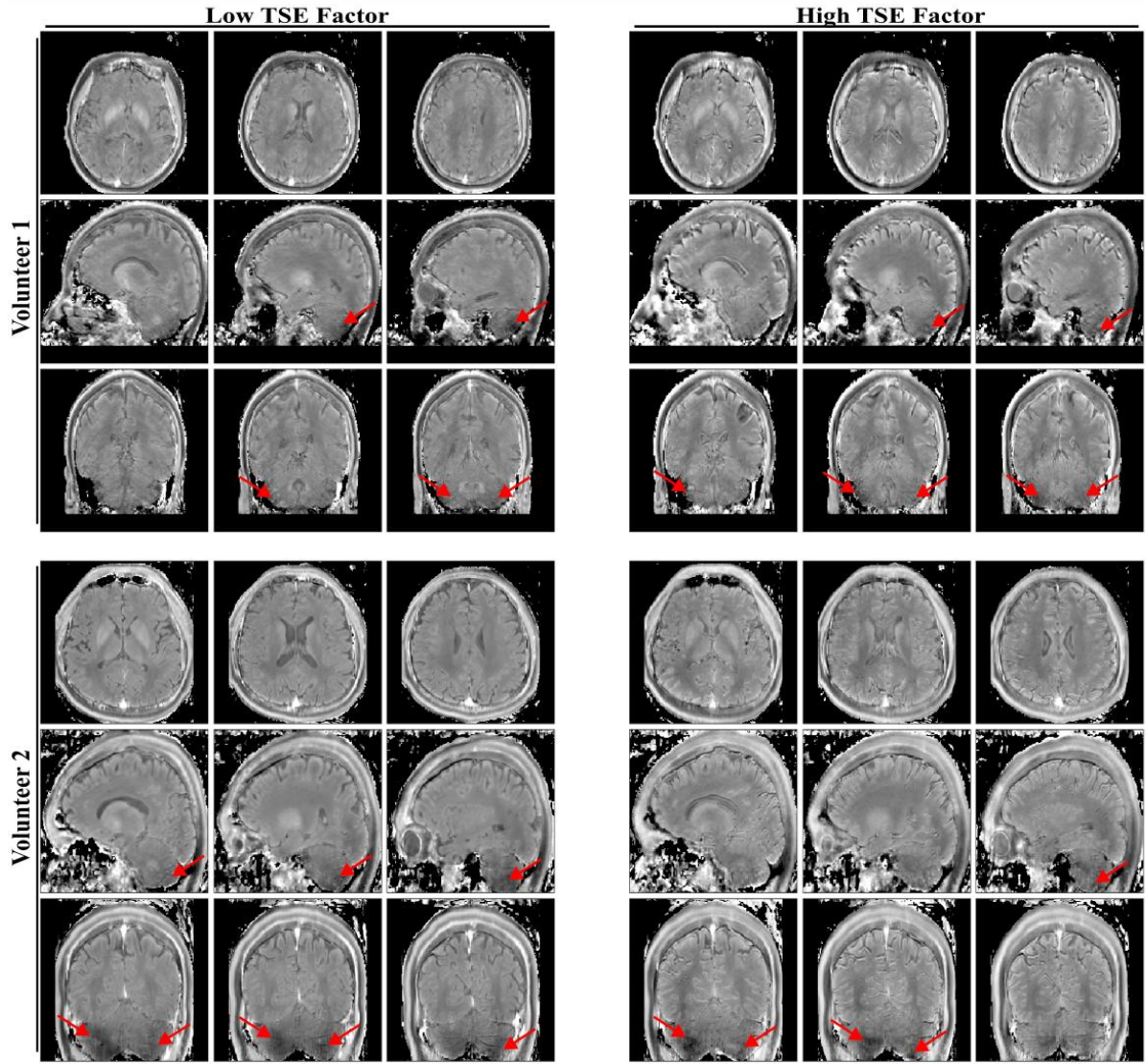

**Figure 7:** Homogeneous images with PD contrast obtained via BOGA method using TSE acquisitions with $SP_1$, $SP_2$ and $SP_3$, in transversal, sagittal and coronal orientations. Dark regions can be observed at the cerebellum and brain stem regions where coil coverage is not sufficient for either RF mode. Dark regions are indicated with red arrows in sagittal and coronal slices.

Figure 8 demonstrates the normalized profiles in x, y, and z direction for $T_1$, $T_2$ and PD weighted images obtained with the BOGA method and CP mode images with respective SPs of the $S_1$ and $S_2$ Profile locations are illustrated on CP mode images for both volunteers acquired with low TSE factors and SP$_1$. Normalized profiles for $T_1$, $T_2$ and PD weighted images are compared against the corresponding SP for generating the same contrast. Profiles of $T_1$ weighted images are compared against profiles of CP mode TSE images obtained with SP$_1$ (short TE and short TR). Similarly, $T_2$ and PD weighted images are compared against profiles of CP mode TSE images obtained with SP$_3$ (long TE and long TR) and SP$_2$ (short TE and long TR). Normalized profiles of the CP mode TSE images show significant spatial variations in signal intensity corresponding the $B_1$ inhomogeneity as expected. On the other hand, profiles for the images obtained via BOGA method do not suffer from such spatial variations, indicating that $B_1$ inhomogeneity are not present in the final images with $T_1$, $T_2$ and PD contrasts. It is also worth noting that CP mode TSE images and images obtained via the BOGA method do not have the same intensity changes due to the tissue profiles as CP mode images have mixed or $T_2$ weighted image contrasts for T$_1$ and PD images which depend on the TSE factor. In addition, the spatial variations in CP mode images are a clear indicator of the $B_1$ inhomogeneity.

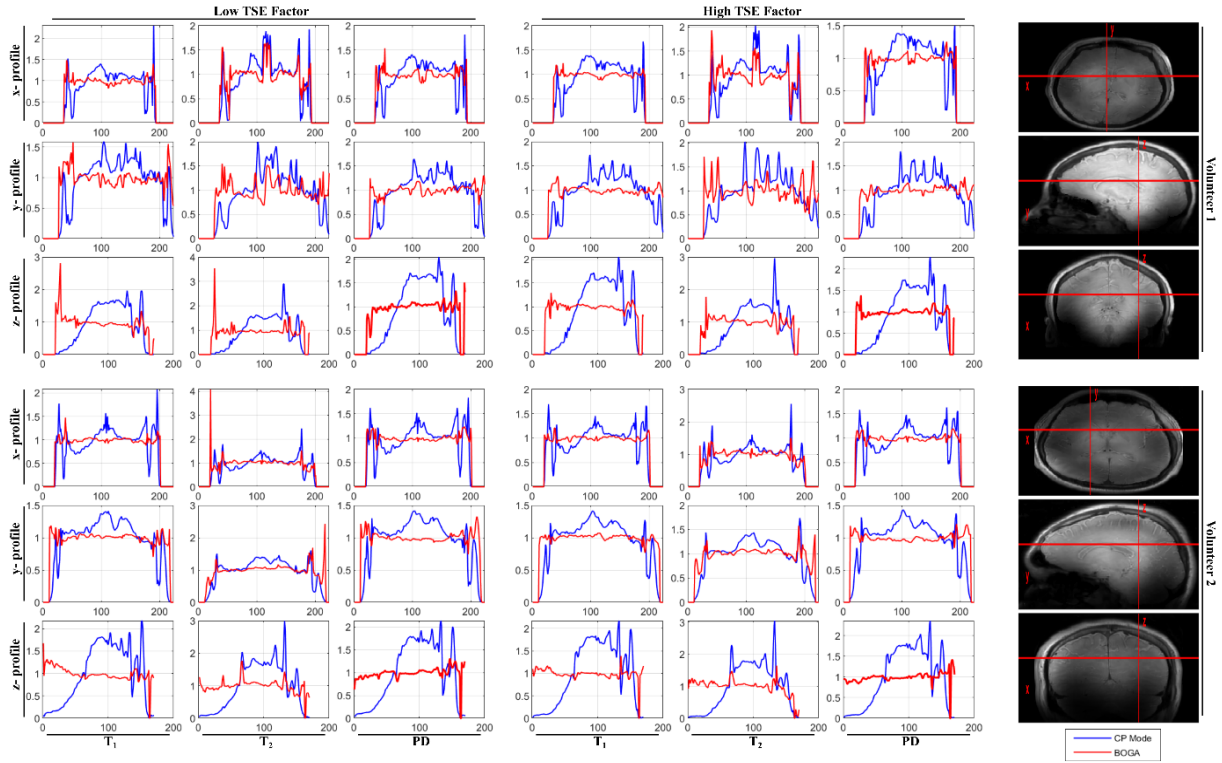

**Figure 8:** Normalized intensity profiles in x, y and z directions for the homogeneous $T_1$, $T_2$ and PD weighted images obtained via BOGA method and CP mode images with corresponding SP. All the profiles for the CP mode images have baseline variations due to the $B_1$ inhomogeneity regardless of the SP and TSE factor. Whereas profiles for the $T_1$, $T_2$ and PD weighted images obtained via BOGA method do not show such baseline variations, further confirming the homogeneity of the images.

SNR maps for the transversal slice for homogeneous $T_1$, $T_2$ and PD weighted images obtained with the BOGA method are illustrated in Figure 9 for low and high TSE factors. SNR maps for CP mode images with SP$_1$, SP$_2$ and SP$_3$ with low and high TSE factors are demonstrated in Figure 10. Mean and standard deviation of SNR for each image in Figure 9 and 10 are provided in Table 2. Considering that the high TSE factor accommodates a longer echo train lengths than the low TSE acquisition, high TSE factor acquisitions can be utilized for faster scans with the BOGA method. Due to the increased TR as a side effect of longer TSE train length BOGA with high TSE factors leads to an increase in overall SNR of $T_1$ and PD weighted images (~34.5% for $T_1$ and ~20% for PD), which is observable at the SNR maps for the high TSE factor images.

It can also be seen from the SNR maps in Figure 9 and 10 and the average SNR values provided in Table 2, that SNR of $T_2$ weighted images obtained via the BOGA method is similar for both low and high TSE factors. The total scan time is 36 minutes 36 seconds and 27 minutes 52 seconds for low TSE versus high TSE factor images, respectively. Thus the increase in SNR of BOGA derived $T_1$ and PD weighted images acquired with high TSE factors is also accompanied by a decrease of the total scan time required for obtaining the $T_1$, $T_2$ and PD contrast. The observed SNR gain is likely a consequence of the increase in TR in the high TSE factor case, whereas the utilized TE for both low and high TSE factor acquisitions of the BOGA method are identical.

Furthermore, SNR maps in Figure 9 and 10 also show that SNR for $T_1$ and PD weighted images increases with the BOGA method compared to the CP mode whereas SNR for the $T_2$ weighted images decrease. This is further verified by the average SNR values in Table 2, as $T_1$ and PD weighted images have approximately 38% and 34% higher SNR values than the corresponding CP mode images with $SP_1$ and $SP_2$ with the exception of low TSE acquisitions for volunteer 1. In contrast average SNR for $T_2$ weighted images show approximately 17% decrease compared to the CP mode images with $SP_3$. This dependency of changes in SNR to the contrast can be attributed to the removal of additional $T_2$ weighting caused by the TSE acquisitions.

|  | CP Mode | | | BOGA method | | |
|---|---|---|---|---|---|---|
|  | $SP_1$ | $SP_2$ | $SP_3$ | $T_1$ | $T_2$ | PD |
| Volunteer 1 low TSE factor | 45.59±43.8 | 41.22±36.0 | 31.85±27.8 | 36.43±28.9 | 28.92±22.2 | 39.55±29.5 |
| Volunteer 1 high TSE factor | 41.14±41.7 | 39.46±43.0 | 36.97±36.0 | 56.51±40.6 | 29.99±25.3 | 47.93±38.0 |
| Volunteer 2 low TSE factor | 41.27±38.8 | 38.22±36.0 | 31.62±24.5 | 50.72±41.5 | 30.86±25.7 | 47.80±40.3 |
| Volunteer 2 high TSE factor | 38.61±41.2 | 35.89±36.5 | 37.00±35.4 | 60.43±44.8 | 31.56±23.7 | 56.65±41.4 |

**Table 2:** Mean and standard deviation of the SNR maps shown in Figures 9 and 10 for CP mode images with $SP_1$, $SP_2$ and $SP_3$ and $T_1$, $T_2$ and PD weighted images obtained via BOGA method.

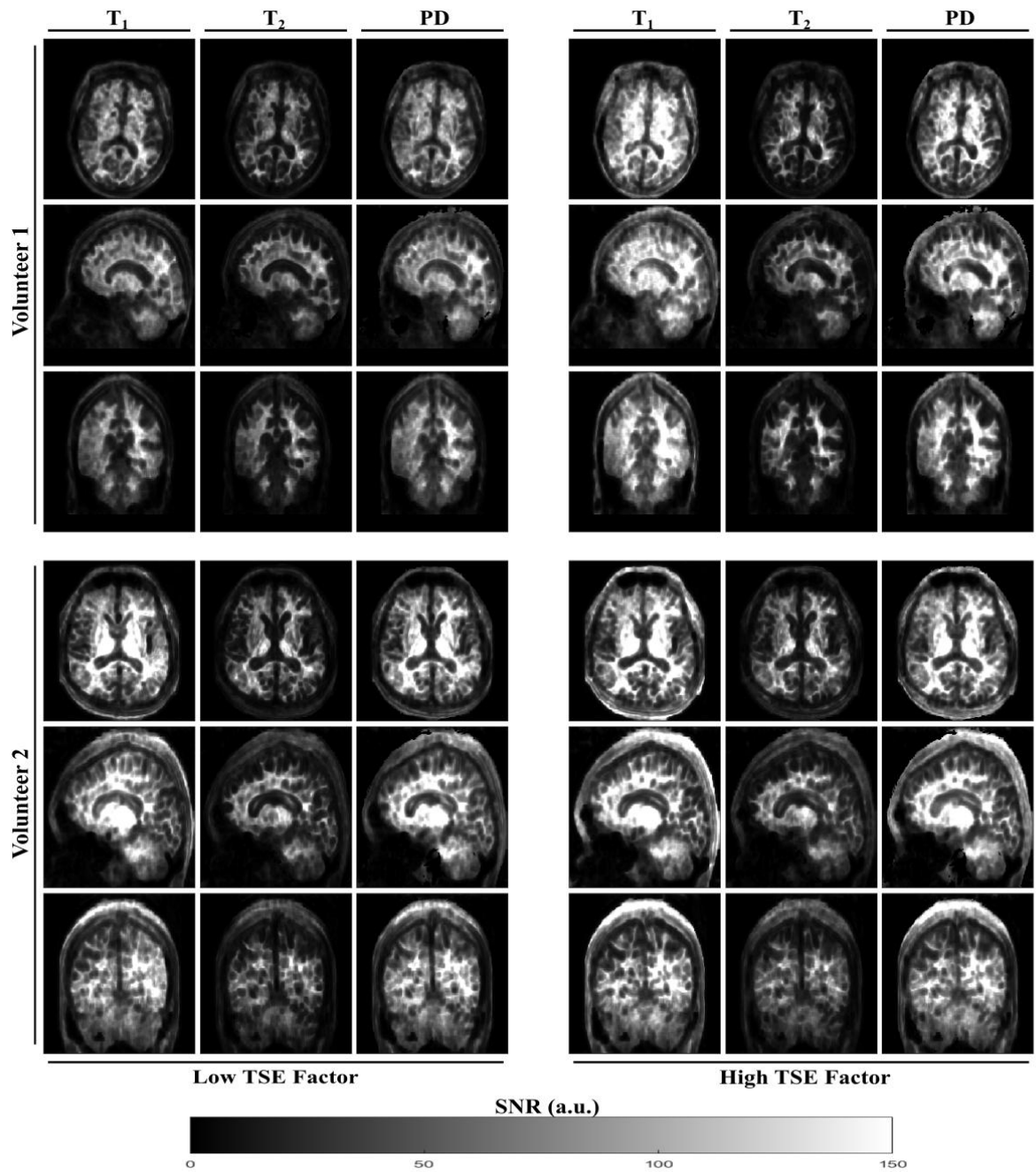

**Figure 9:** SNR maps calculated for the homogeneous $T_1$, $T_2$ and PD weighted images obtained via BOGA method. Overall increase SNR in $T_1$ and PD weighted images with respect to the increase in TSE factor can be observed for each homogeneous image obtained via BOGA method.

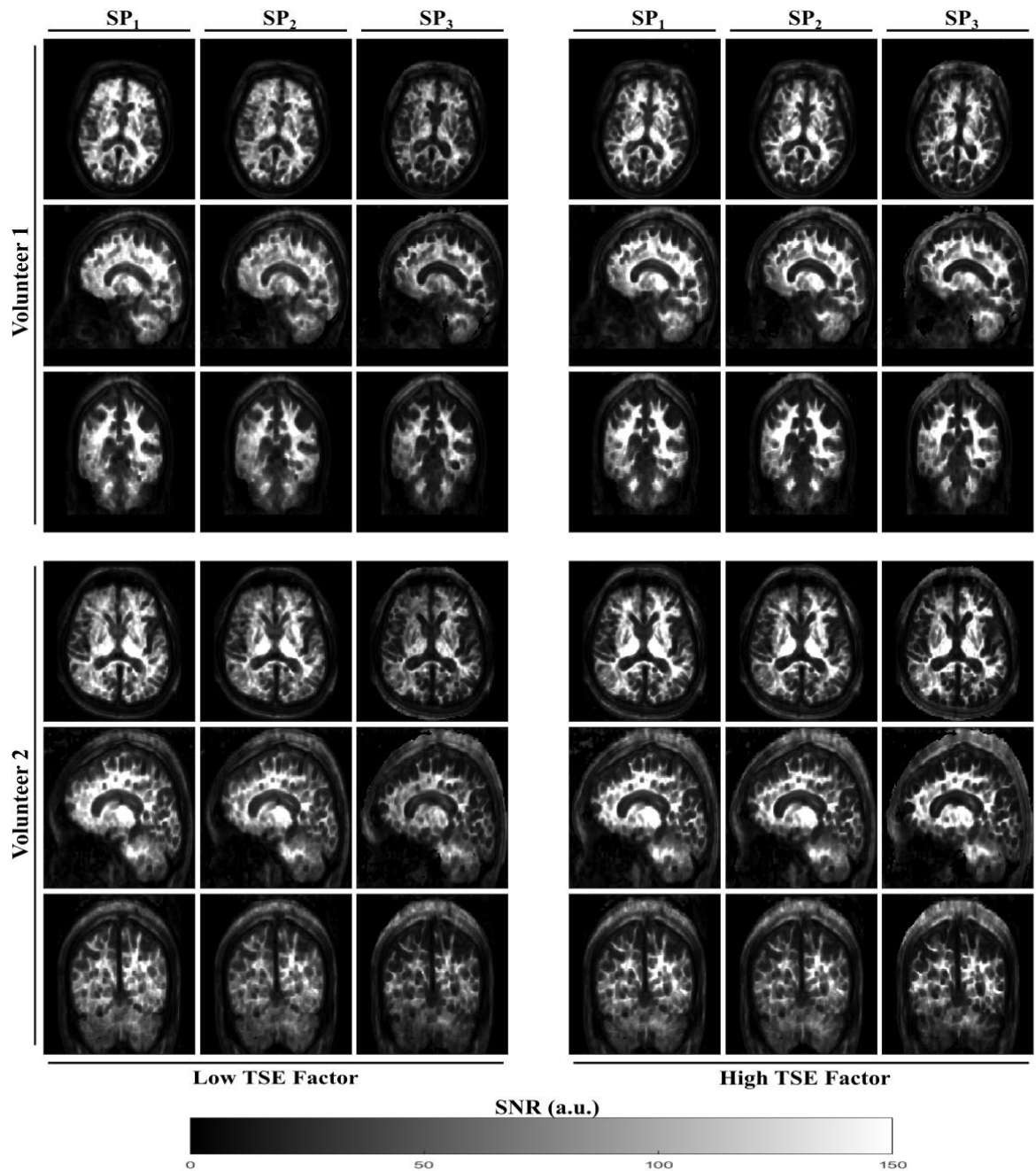

**Figure 10:** SNR maps calculated for the CP mode images with $SP_1$, $SP_2$ and $SP_3$. Overall increase SNR related to increase in TR can be observed for both volunteers with the high TSE factor.

## Discussion

It has been shown that homogeneous whole brain images with $T_1$, $T_2$ and PD contrast at 7T can be obtained without $B_1$ inhomogeneity effects via the BOGA method adopted to TSE imaging, as shown in Figure 5, 6 and 7. TSE images obtained using the regular CP operating mode, of the scanner, are affected from $B_1$ inhomogeneity and have either mixed or dominant $T_2$ contrast (Figure 1). Since the virtual transmit channel images are calculated individually for the application of the BOGA method in

TSE imaging and effects of $T_2$ weighting due to the spin echo train length is constant for each acquisition, both contrast change and $B_1$ inhomogeneity issues are solved simultaneously by the BOGA method. Since the contrast change is treated as a part of the $B_1$ inhomogeneity, effect of the TSE factor can be mitigated. This is clearly observable in the $T_1$, $T_2$ and PD weighted whole brain images (Figure 5, 6, and 7) obtained with the BOGA method as they show homogeneous and clear image contrast unlike the CP mode images demonstrated in Figure 1.

In Figures 5, 6 and 7, removal of the $B_1$ inhomogeneity has been demonstrated. However, low signal intensity of the dual channel transmit coil at the cerebellum limits the coverage of the region, resulting in high and low intensity regions in $T_1$ and PD weighted images. As it is a limitation imposed by the RF coil itself, this issue can be solved by the utilization of the 8-channel transmit coil version of the same coil vendor with similar channel groupings as in[40]. It has been previously shown that with the respective 8-channel transmit RF coil good coverage of the cerebellum can be achieved with the BOGA method. Furthermore, more complex RF modes can also be utilized if more transmit channels become available for guaranteed non overlapping low intensity regions as the exact $B_1$ inhomogeneity patterns are irrelevant to the BOGA method.

TSE images obtained for each SP and RF mode are used twice in the calculation of respective $S_3$ and $S_4$ images and once as the $S_1$ and $S_2$ depending on the RF mode for application of the BOGA mode. This demonstrates the full utilization of the total scan time as all six acquired data sets are used multiple times for simultaneously obtaining homogeneous $T_1$, $T_2$ and PD contrast without any prior measurement scans and/or RF shim/pulse design. However, it is worth noting that the scan time efficiency of the BOGA method will decrease for any single image contrast, because of the requirement for four data acquisitions with two different respective SPs and RF modes for obtaining any single contrast. In contrast in the multiple image contrast case, all three contrasts ($T_1$, $T_2$ and PD) can be obtained with six data acquisitions. This indicates that multi contrast TSE application of the BOGA method is more scan time efficient and hence preferable compared to the single contrast case.

Since contrast in the images obtained with the BOGA method depends on the differences between SPs, these selections of SPs enable realization of the BOGA method for $T_1$, $T_2$ and PD, independently of the image contrast of the initial images. As it has been demonstrated with input images with mixed contrast for the low TSE factor and with input images with $T_2$ contrast for high TSE factor the BOGA method enables pure $T_1$, $T_2$ and PD image contrasts. However, caution must be exercised for determining SPs and TSE factors since strong $T_2$ weighting that is common in all SPs can diminish the $T_1$ and PD contrast similarly to a signal dropout, resulting in artificial lack of coil coverage, which cannot be mitigated by the BOGA method. Which can suggest a limit on the TSE factor depending on the target tissues properties. An example is shown for volunteer 2, as hyperintense regions in cerebellum are present for high TSE factor $T_2$ images and not present in low TSE factor $T_2$ images. However, as demonstrated herein a high TSE factor of 100 can be utilized for obtaining homogeneous $T_1$, $T_2$ and PD weighted images from the whole brain with increased SNR in $T_1$ and PD weighted images using the BOGA TSE method. The reason for the increase in the SNR with higher TSE factors is that the increase in TR due to the longer echo train causes less saturation effects which overpowers the decrease in SNR due to $T_2$ decay. This benefit comes along with decreased scan time.

In order to shorten the total scan time, Compressed SENSE acceleration is used along with the TSE acquisition. However, even with the double acceleration of the data acquisition, total scan time is prolonged due to the multiple acquisitions required for all three contrasts. This also leads to a higher vulnerability of BOGA method to subject motion in between scans. Since all the TSE acquisitions are used for obtaining $T_1$, $T_2$ and PD contrasts in the final images, artifacts due to the motion in one image can propagate to the others as well. For reducing the susceptibility of the final images to the motion, integration of the motion correction algorithms in the data acquisition will be investigated in future work. Furthermore, reducing the effect of motion and further reducing the total scan time, integration of time interleaving for the data acquisitions, similar to TIAMO, can also be explored[19-22].

# Conclusion

The BOGA method that has been previously introduced for homogeneous $T_2^*$ contrast for whole brain, is further extended for the application in TSE imaging to obtain homogeneous $T_1$, $T_2$ and PD contrasts for whole brain imaging without $B_1$ inhomogeneity effects. Final homogeneous images are obtained through the combination of the input TSE images acquired with three distinct SPs and two RF modes. Furthermore, mixed contrast effects or T2 weighting caused by the choice of the TSE factor are simultaneously eliminated with the $B_1$ inhomogeneity effects.

Compared to the TSE images obtained with the CP mode, $T_1$, $T_2$ and PD weighted images obtained with the BOGA methods are free of $B_1$ inhomogeneity effects and have consistent image contrast. Moreover, it is also shown that increased $T_2$ weighting due to the high TSE factors are also resolved with the BOGA method indicating that both the $B_1$ inhomogeneity and the inconsistent contrast issues of the TSE imaging at 7T can be addressed by this method. In the future, application in 8 channel system and integration of higher acceleration factors will be investigated for achieving better coverage in a shorter scan time as well as prospective motion correction.

**References:**

1. Hennig J, Nauerth A, Friedburg H. RARE imaging: A fast imaging method for clinical MR. *Magnetic Resonance in Medicine*. 1986;3(6):823-833. doi:https://doi.org/10.1002/mrm.1910030602

2. MRI Sequences: acronyms | e-MRI. IMAIOS. https://www.imaios.com/en/e-mri/sequences/sequences-acronyms

3. Hoch MJ, Bruno MT, Faustin A, et al. 3T MRI Whole-Brain Microscopy Discrimination of Subcortical Anatomy, Part 1: Brain Stem. *American Journal of Neuroradiology*. Published online January 31, 2019. doi:https://doi.org/10.3174/ajnr.a5956

4. Priovoulos N, Jacobs HIL, Ivanov D, Uludağ K, Verhey FRJ, Poser BA. High-resolution in vivo imaging of human locus coeruleus by magnetization transfer MRI at 3T and 7T. *NeuroImage*. 2018;168:427-436. doi:https://doi.org/10.1016/j.neuroimage.2017.07.045

5. L X. T2 Relaxation Time Mapping of the Lumbar Spine at 3T MRI -Different Protocols Comparison. *Clinical Radiology & Imaging Journal*. 2020;4(1). doi:https://doi.org/10.23880/crij-16000165

6. Bot JCJ, Barkhof F. Spinal-Cord MRI in Multiple Sclerosis: Conventional and Nonconventional MR Techniques. *Neuroimaging Clinics of North America*. 2009;19(1):81-99. doi:https://doi.org/10.1016/j.nic.2008.09.005

7. Garibaldi M, Tasca G, Jordi Diaz-Manera, et al. Muscle MRI in neutral lipid storage disease (NLSD). *Journal of Neurology*. 2017;264(7):1334-1342. doi:https://doi.org/10.1007/s00415-017-8498-8

8. Mellado JM, Pérez del Palomar L. Muscle hernias of the lower leg: MRI findings. *Skeletal Radiology*. 1999;28(8):465-469. doi:https://doi.org/10.1007/s002560050548

9. Tona KD, van Osch MJP, Nieuwenhuis S, Keuken MC. Quantifying the contrast of the human locus coeruleus in vivo at 7 Tesla MRI. Todd N, ed. *PLOS ONE*. 2019;14(2):e0209842. doi:https://doi.org/10.1371/journal.pone.0209842

10. Madai VI, von Samson-Himmelstjerna FC, Bauer M, et al. Ultrahigh-Field MRI in Human Ischemic Stroke – a 7 Tesla Study. *PLoS ONE*. 2012;7(5). doi:https://doi.org/10.1371/journal.pone.0037631

11. Laistler E, Dymerska B, Sieg J, et al. In vivo MRI of the human finger at 7 T. *Magnetic Resonance in Medicine*. 2017;79(1):588-592. doi:https://doi.org/10.1002/mrm.26645

12. Kraff O, Quick HH. 7T: Physics, safety, and potential clinical applications. *Journal of Magnetic Resonance Imaging*. 2017;46(6):1573-1589. doi:https://doi.org/10.1002/jmri.25723

13. Bae KT, Park SH, Moon CH, Kim JH, Kaya D, Zhao T. Dual-echo arteriovenography imaging with 7T MRI. Journal of Magnetic Resonance Imaging. 2009;31(1):255-261. doi:https://doi.org/10.1002/jmri.22019

14. Mistry N, Tallantyre EC, Dixon JE, et al. Focal multiple sclerosis lesions abound in "normal appearing white matter." *Multiple Sclerosis Journal*. 2011;17(11):1313-1323. doi:https://doi.org/10.1177/1352458511415305

15. Barisano G, Sepehrband F, Ma S, et al. Clinical 7 T MRI: Are we there yet? A review about magnetic resonance imaging at ultra-high field. *The British Journal of Radiology*. 2019;92(1094):20180492. doi:https://doi.org/10.1259/bjr.20180492

16. Vargas MI, Martelli P, Xin L, et al. Clinical Neuroimaging Using 7 T MRI: Challenges and Prospects. *Journal of Neuroimaging*. 2017;28(1):5-13. doi:https://doi.org/10.1111/jon.12481

17. Pazahr S, Nanz D, Sutter R. 7 T Musculoskeletal MRI. *Investigative Radiology*. 2022;Publish Ahead of Print. doi:https://doi.org/10.1097/rli.0000000000000896

18. van Kalleveen IML, Koning W, Boer VO, Luijten PR, Zwanenburg JJM, Klomp DWJ. Adiabatic turbo spin echo in human applications at 7 T. *Magnetic Resonance in Medicine*. 2011;68(2):580-587. doi:https://doi.org/10.1002/mrm.23264

19. Orzada S, Maderwald S, Poser BA, Bitz AK, Quick HH, Ladd ME. RF excitation using time interleaved acquisition of modes (TIAMO) to address B 1 inhomogeneity in high-field MRI. *Magnetic Resonance in Medicine*. 2010;64(2):327-333. doi:https://doi.org/10.1002/mrm.22527

20. He X, Schmidt S, Štefan Zbýň, Haluptzok T, Moeller S, Metzger GJ. Improved TSE imaging at ultrahigh field using nonlocalized efficiency RF shimming and acquisition modes optimized for refocused echoes (AMORE). *Magnetic Resonance in Medicine*. 2022;88(4):1702-1719. doi:https://doi.org/10.1002/mrm.29318

21. Sascha Brunheim, Gratz M, Sören Johst, et al. Fast and accurate multi-channel mapping based on the TIAMO technique for 7T UHF body MRI. *Magnetic Resonance in Medicine*. 2018;79(5):2652-2664. doi:https://doi.org/10.1002/mrm.26925

22. Orzada S, Johst S, Maderwald S, Bitz AK, Solbach K, Ladd ME. Mitigation of $B_1^+$ inhomogeneity on single-channel transmit systems with TIAMO. *Magnetic Resonance in Medicine*. 2012;70(1):290-294. doi:https://doi.org/10.1002/mrm.24453

23. Suwit Saekho, Boada FE, Noll DC, V. Andrew Stenger. Small tip angle three-dimensional tailored radiofrequency slab-select pulse for reduced $B_1$ inhomogeneity at 3 T. *Magnetic Resonance in Medicine*. 2005;53(2):479-484. doi:https://doi.org/10.1002/mrm.20358

24. Grissom W, Yip C, Zhang Z, Stenger VA, Fessler JA, Noll DC. Spatial domain method for the design of RF pulses in multicoil parallel excitation. *Magnetic Resonance in Medicine*. 2006;56(3):620-629. doi:https://doi.org/10.1002/mrm.20978

25. Grissom WA, Yip CY, Wright SM, Fessler JA, Noll DC. Additive angle method for fast large-tip-angle RF pulse design in parallel excitation. *Magnetic Resonance in Medicine*. 2008;59(4):779-787. doi:https://doi.org/10.1002/mrm.21510

26. Schmitter S, DelaBarre L, Wu X, et al. Cardiac imaging at 7 tesla: Single- and two-spoke radiofrequency pulse design with 16-channel parallel excitation. 2013;70(5):1210-1219. doi:https://doi.org/10.1002/mrm.24935

27. Cloos MA, Boulant N, Luong M, et al. kT-points: Short three-dimensional tailored RF pulses for flip-angle homogenization over an extended volume. *Magnetic Resonance in Medicine*. 2011;67(1):72-80. doi:https://doi.org/10.1002/mrm.22978

28. Xu D, King KF, Zhu Y, McKinnon GC, Liang ZP. Designing multichannel, multidimensional, arbitrary flip angle RF pulses using an optimal control approach. *Magnetic Resonance in Medicine*. 2008;59(3):547-560. doi:https://doi.org/10.1002/mrm.21485

29. Pohmann R, Scheffler K. A theoretical and experimental comparison of different techniques for $B_1$ mapping at very high fields. *NMR in Biomedicine*. 2012;26(3):265-275. doi:https://doi.org/10.1002/nbm.2844

30. Gras V, Vignaud A, Amadon A, Bihan D, Boulant N. Universal pulses: A new concept for calibration-free parallel transmission. *Magnetic Resonance in Medicine*. 2016;77(2):635-643. doi:https://doi.org/10.1002/mrm.26148

31. Geldschläger O, Bosch D, Glaser S, Henning A. Local excitation universal parallel transmit pulses at 9.4T. *Magnetic Resonance in Medicine*. 2021;86(5):2589-2603. doi:https://doi.org/10.1002/mrm.28905

32. Gras V, Mauconduit F, Vignaud A, et al. Design of universal parallel-transmit refocusing kT -point pulses and application to 3D T2 -weighted imaging at 7T. *Magnetic Resonance in Medicine*. 2017;80(1):53-65. doi:https://doi.org/10.1002/mrm.27001

33. Gras V, Pracht ED, Mauconduit F, Le Bihan D, Stöcker T, Boulant N. Robust nonadiabatic T2 preparation using universal parallel-transmit kT -point pulses for 3D FLAIR imaging at 7 T. *Magnetic Resonance in Medicine*. 2019;81(5):3202-3208. doi:https://doi.org/10.1002/mrm.27645

34. Herrler J, Liebig P, Gumbrecht R, et al. Fast online-customized (FOCUS) parallel transmission pulses: A combination of universal pulses and individual optimization. *Magnetic Resonance in Medicine*. 2021;85(6):3140-3153. doi:https://doi.org/10.1002/mrm.28643

35. Busse RF, Hariharan H, Vu A, Brittain JH. Fast spin echo sequences with very long echo trains: Design of variable refocusing flip angle schedules and generation of clinical T2 contrast. *Magnetic Resonance in Medicine*. 2006;55(5):1030-1037. doi:https://doi.org/10.1002/mrm.20863

36. Ladd ME. High-Field-Strength Magnetic Resonance. *Topics in Magnetic Resonance Imaging*. 2007;18(2):139-152. doi:https://doi.org/10.1097/rmr.0b013e3180f612b3

37. Michaeli S, Garwood M, Zhu X, et al. Proton $T_2$ relaxation study of water, N-acetylaspartate, and creatine in human brain using Hahn and Carr-Purcell spin echoes at 4T and 7T. *Magnetic Resonance in Medicine*. 2002;47(4):629-633. doi:https://doi.org/10.1002/mrm.10135

38. Boğa Ç, Henning A. Bilateral Orthogonality Generating Acquisitions Method for Homogeneous $T_2^*$ Images Using Dual Channel Parallel Transmission at 7T. In Proceedings of the 31th Annual Meeting of ISMRM, Toronto, Canada, 2023

39. Boğa Ç, Henning A. Simultaneous $R_2^*$ Relaxometry and Dixon Imaging of Liver and Kidney at 3T Using Bilateral Orthogonality Generative Acquisitions Method. In Proceedings of the 32th Annual Meeting of ISMRM, Singapore, 2024

40. Boğa Ç, Henning A. Bilateral orthogonality generative acquisitions method for homogeneous T2* images using parallel transmission at 7 T. *Magnetic Resonance in Medicine*. Published online October 7, 2024. doi:https://doi.org/10.1002/mrm.30329 '

41. McCann AJ, Workman A, McGrath C. A quick and robust method for measurement of signal-to-noise ratio in MRI. *Physics in Medicine and Biology*. 2013;58(11):3775-3790. doi:https://doi.org/10.1088/0031-9155/58/11/3775

42. movingstd & movingstd2 - File Exchange - MATLAB Central. www.mathworks.com. https://www.mathworks.com/matlabcentral/fileexchange/9428-movingstd-movingstd2